\documentclass[a4paper]{jpconf}
\usepackage{graphicx,psfig,epsfig}
\newcommand{\be}{\begin{eqnarray}}
\newcommand{\ee}{\end{eqnarray}}

\newcommand{\ave}[1]{\left\langle #1 \right\rangle}

 \newcommand{\Nch}{N_{\rm ch}}
\begin{document}
\title{Constraining freeze-out with yields and fluctuations}
\author{Giorgio Torrieri}
\address{
Department of Physics, McGill University, Montreal, QC H3A-2T8, Canada}
\author{Sangyong Jeon}
\address{
Department of Physics, McGill University, Montreal, QC H3A-2T8, Canada and RIKEN-BNL Research Center, Upton NY, 11973, USA}
\author{Johann Rafelski}
\address{Department of Physics, University of Arizona, Tucson, Arizona, 85721, USA}
\date{September, 2005}

\begin{abstract}
We show that the simultaneous measurement of yields and fluctuations in heavy ion collisions is capable of falsifying and constraining the statistical hadronization model.   We show how such a measurement can test for chemical non-equilibrium, and distinguish between a high temperature chemically equilibrated freeze-out from a supercooled freeze-out with an over-saturated phase space.   We further explain how this measurement can be used to obtain a model-independent 
estimate of the difference between ``chemical'' and ``thermal'' freeze-out.
\end{abstract}
%\pacs{25.75.-q,24.60.-k,24.10.Pa}
%%%%%%%%%%%%%%%%%%%%%%%%%%%%%%%%%%%%%%%%%%%%

%\maketitle
The statistical
hadronization model (SHM) 
\cite{Fer50,Pom51,Lan53} has been extensively applied to the study of soft particle production in hadronic systems.   When it includes the full resonance spectrum \cite{Hag65}, the SHM can describe quantitatively
the yields of all hadrons produced in heavy ion collisions.   

The ability of the SHM to describe not just averages, but event-by-event multiplicity fluctuations
has however not been widely investigated, and its applicability is currently a matter of controversy.
Event-by-event fluctuations are
 subject to  intense current
theoretical~\cite{fluct1},
and experimental  interest~\cite{starfluct2,phefluct}, as a constraint
for existing models and as a signature of new physics.

This study illustrates the use of both yields and fluctuations as a probe capable of constraining the SHM and differentiating between freeze-out scenarios.

The statistical hadronization model assumes that particles are produced according to a probability determined by their phase
space density.   The first and second cumulants of this probability distribution give, respectively, the average value (over all events) of the desired observable, and its event-by-event fluctuation.

Conserved quantities can be treated in several ways, appropriate to different experimental situations:  If the totality of the system is observed, than conserved quantities
can not fluctuate.   If a small fraction equilibrated with the rest of the system is observed, than conserved quantities will fluctuate event-by-event.
Rigorous conservation is known as the Microcanonical ensemble, while allowing energy and other conserved quantities to fluctuate between the system and the bath leads, respectively, to the Canonical and Grand Canonical (GC) ensembles.    \textit{All} fluctuations are ensemble-specific even in the thermodynamic limit \cite{nogc2}.

 In this work, we use the GC ensemble, implemented in open-source software \cite{share} to calculate fluctuations and yields.  We motivate this choice by the fact that RHIC experiments
observe the mid-rapidity slice of the system, comprising roughly 1/8 of the total multiplicity, an appropriate fraction for a GC prescription.  Boost invariance, a good symmetry around mid-rapidity, links this rapidity slice with a sector in configuration space.
If the system observed at RHIC is a nearly ideal fluid, the matter created in this space should be in equilibrium, grand-canonically, with the unobserved regions.
    If freeze-out temperature throughout observed space is approximately constant, the GC ensemble should be able to describe both yields and fluctuations \cite{cleymans,prlfluct}.

The final state yield of particle can then be computed as a function of the particle mass and resonance decay tree, as well as temperature and chemical potentials.  (The technical details of the calculation are given in a parallel paper \cite{qm2005nukleonika}).
The chemical potential,implemented here via the fugacity $\lambda=e^{\mu/T}$, is the main distinguishing feature between competing freeze-out models.  Provided the law of mass action holds, it should be given by the product of charge fugacities (flavor, isospin etc.).  It is then convenient to parametrize it in terms of equilibrium fugacities $\lambda^{\mathrm{eq}}$ and phase space occupancies $\gamma$.   For a hadron $i$ with $q (\overline{q})$ light quarks, $s (\overline{s})$ strange quarks and isospin $I_3$ the fugacity is then
\begin{eqnarray}
\label{chemneq}
\lambda_i = \lambda_i^{\mathrm{eq}}
\gamma_q^{q+\overline{q}} \gamma_s^{s+\overline{s}}
\phantom{A},\phantom{A}
\lambda_i^{\mathrm{eq}}=\lambda_{q}^{q-\overline{q}} \lambda_{s}^{s -
\overline{s}}
\lambda_{I_3}^{I_3} 
\end{eqnarray}
If the system is in chemical equilibrium then detailed balance requires that $\gamma_q=\gamma_s=1$.   In an expanding system, however, the condition of chemical equilibrium 
no longer holds.  Kinetically, this occurs because collective expansion and cooling will make it impossible for endothermic and exothermic reactions, or for creation and destruction
reactions of a rare particle, to be balanced. Provided the system remains in local thermal equilibrium, $\lambda_i$ can still be used as a Lagrange multiplier for the
particle density, and the first and second cumulants can be calculated from the partition function the usual way \cite{jansbook,chemistrywebsite}.  However, in this case in general
$\gamma_q \ne 1,\gamma_s \ne 1$.

This picture becomes particularly appropriate if the expanding system undergoes a fast phase transition from a QGP to a hadron gas (HG).  In this case, chemical non-equilibrium  \cite{sudden} and super-cooling \cite{csorgo} can arise due to entropy conservation:
By dropping the hadronization temperature to $\sim 140 MeV$ and oversaturating the phase space above equilibrium ($\gamma_q \sim 1.5,\gamma_s \sim 2$), it is possible to match the entropy of a HG with that of a QGP \cite{sudden}.

Fits to experimental data  at both SPS
and RHIC energies  indeed support these values of $\gamma_{q,s}$ when these parameters are fitted.  Moreover, best fit $\gamma_{q,s}>1$ arises for a critical energy \cite{gammaq_energy}  (corresponding to the energy of the $K/\pi$ ``horn'' \cite{horn}) and system size \cite{gammaq_size}, as expected from the interpretation of $\gamma_q$ as a manifestation of a phase transition.  However, the fits performed in \cite{gammaq_energy} have not been able to rule out  equilibrium models (at SPS and RHIC the difference in statistical significance between equilibrium and non-equilibrium is $\sim 20 \%$), which are usually preferred for their 
smaller number of parameters to fit \cite{barannikova,bdm}.    Equilibrium freeze-out temperature varies between fits, ranging from 155 \cite{gammaq_energy} to 177 \cite{bdm} MeV.

Both scenarios are physically reasonable, can describe the data, and would be instrumental in our understanding of hadronic matter if proven correct.   In particular, the HBT puzzle suggests we lack understanding of the last stages of the fireball evolution.    Non-equilibrium is useful in this respect, since it affects both system volume \cite{bari,gammaq_energy} and emission time \cite{csorgo}.

The reason both equilibrium and non-equilibrium are compatible with data is that in a fit to yields the non-equilibrium phase space occupancies
$\gamma_s$ and $\gamma_q$ correlate with freeze-out temperature  \cite{gammaq_energy}, making a distinction between
 a $T=170$ MeV equilibrated freeze-out ($\gamma_q=1,\gamma_s\leq 1$) scenario and a supercooled scenario where $\gamma_{q,s}>1$ problematic.  

A related ambiguity is the difference between chemical freeze-out (where particle abundances are fixed) and thermal
freeze-out (where particles cease to interact).   Equilibrium models generally assume a long phase between these two points, which would alter considerably the multiplicity of directly detectable resonances.  In a Non-equilibrium supercooled freeze-out, on the other
hand, it is natural to assume that particle interaction after emission is negligible \cite{sudden}.   Once again, a reliable way to probe the extent of the
reinteraction would be instrumental for our understanding of how the fireball produced in heavy ion collisions breaks up.

We have recently shown \cite{prlfluct} that event-by-event fluctuations can be used to solve the dilemmas discussed above.
The equations in \cite{qm2005nukleonika} make it clear that the dependence of fluctuations on $T$ and $\gamma_q$ is different, allowing us to decouple these two variables.  a higher temperature tends to decrease fluctuations w.r.t. the Poisson value expected from Boltzmann statistics
, since it introduces greater correlations due to an increased
resonance contribution.  Increasing $\gamma_q$ will rapidly increase fluctuations of quantities related to pions, due to the fact
that at $\gamma_q>1$ $\lambda_{\pi}$ rapidly approaches $e^{m_{\pi}/T}$, giving fluctuations an extra boost w.r.t. yields \cite{prlfluct,qm2005nukleonika}.

%%%%%%%%%%%%%%%%%%%%%%%%%%
\begin{figure}[h]
%\begin{center}
\psfig{width=5.9cm,clip=,figure=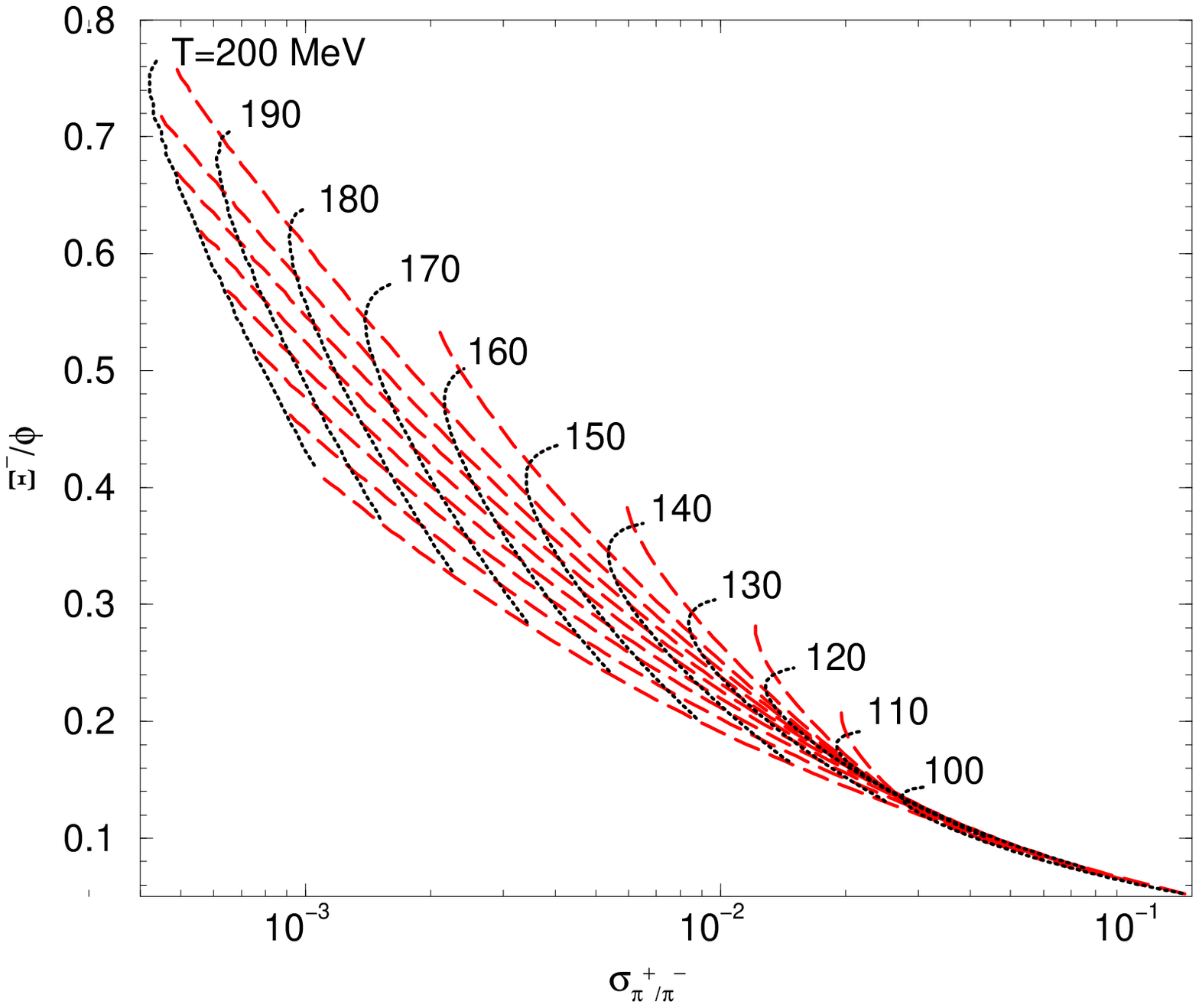}
\psfig{width=5.3cm,clip=,figure=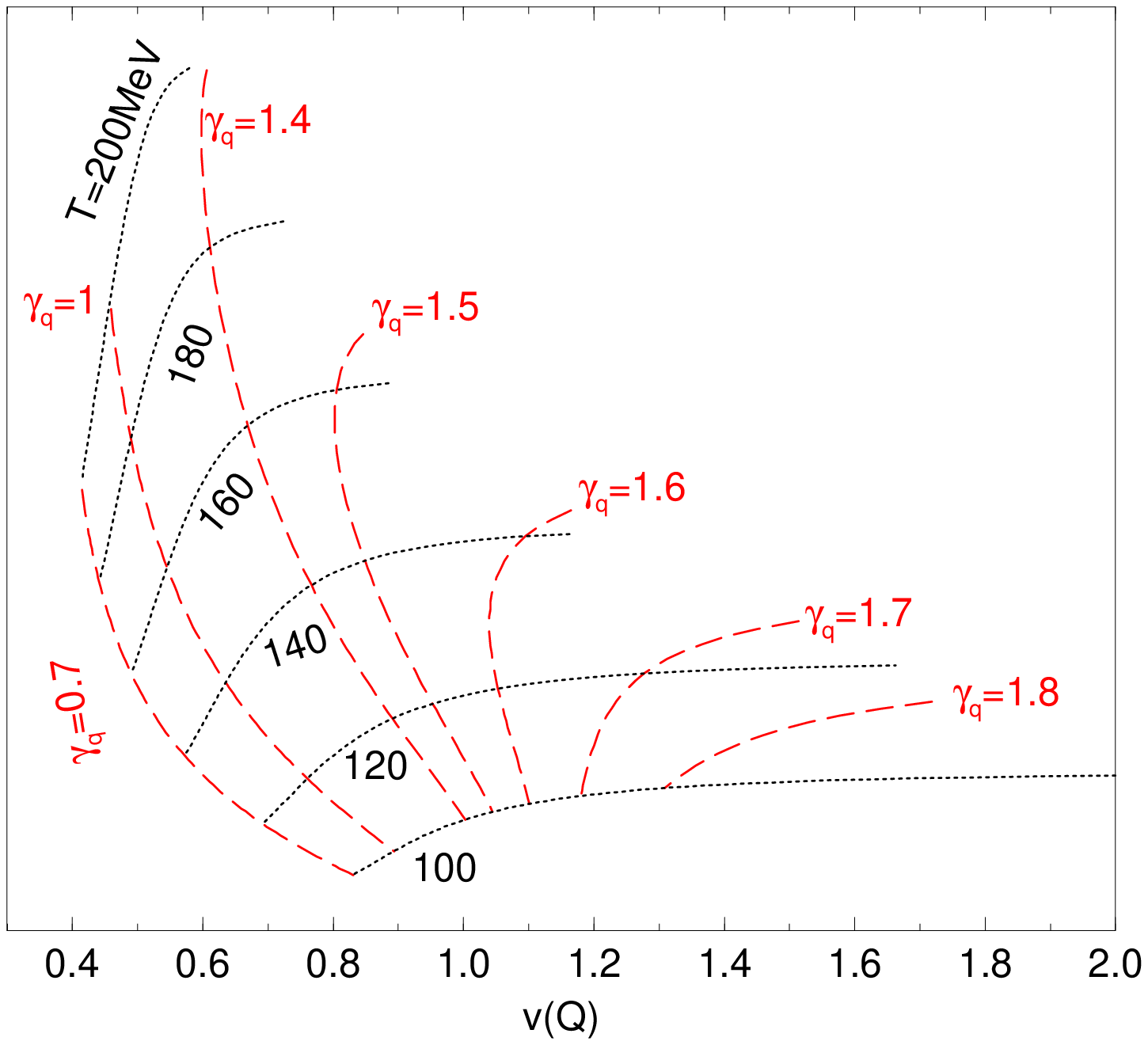}
\psfig{width=5.9cm,clip=,figure=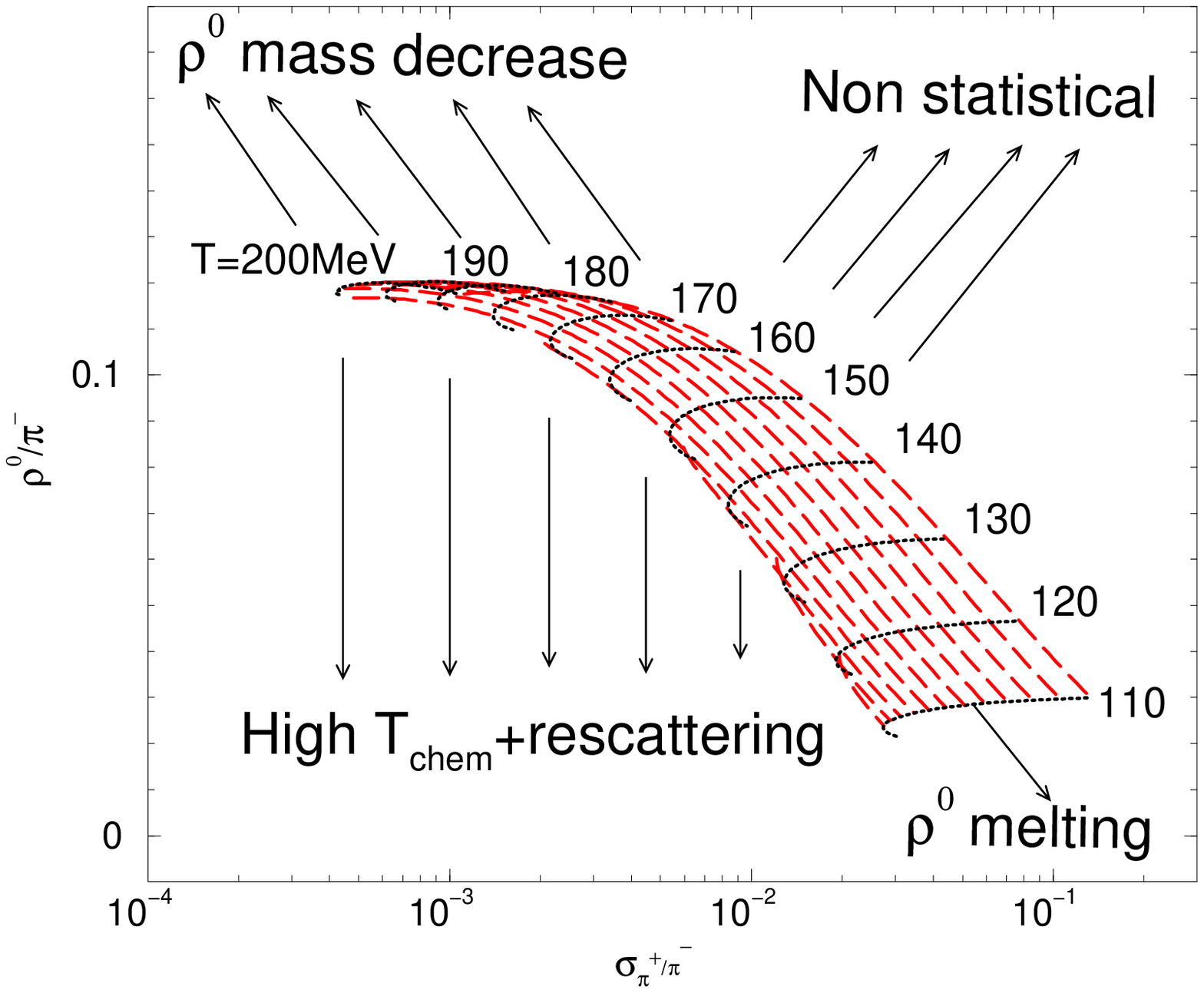}
\caption{\label{xifold} (color online)
Comparing a fluctuation and a particle yield sensitive to $T$ and $\gamma_q$.  Dotted black lines refer equal temperature (T=100-200 MeV), while long-dashed red lines refer to $\gamma_q$ ($\gamma_q=0.7-1.8$).
  Left: $\Xi/\phi$ vs. $\sigma_{\pi^-/\pi^+}$.  Center:  $\Xi/\phi$ vs. $v(Q)$.
Right:   $\rho^0/\pi^-$ vs. $\sigma_{\pi^-/\pi^+}$, which also probes the extent of hadronic interactions after chemical freeze-out.  As shown in \cite{prlfluct}, parameters other than $T$ and $\gamma_q$ do not
impact the observed quantities significantly}
%\end{center}
\end{figure}
%%%%%%%%%%%%%%%%%%%%%

In \cite{prlfluct} we used 
a measurement of the charge fluctuation measure 
$v(Q)=\ave{\Delta Q^2}/\ave{\Nch}$
%\ee
where $\ave{\Delta Q^2}$ is the fluctuation in net charge and $\Nch$ is the charged particle multiplicity.
For the ratio, we used $\Lambda/K^-$, corrected for $\Xi$ and $\phi$ feed-down.

Fig. \ref{xifold} applies this procedure again, this time with the easier to measure $\Xi/\phi$ ratio, which also depends mainly on $T$ and $\gamma_q$ at RHIC (and on $\lambda_s$ at SPS).
The central panel shows $\Xi/\phi$ plotted against $v(Q)$,and the left  panel shows $\Xi/\phi$ plotted against the event-by-event fluctuation in the $\pi^+/\pi^-$ ratio (labeled as $\sigma_{\pi^+/\pi^-}$).  
As the left panel shows, $\gamma_q$ correlates the $\pi^+/\pi^-$ fluctuation with $\Xi/\phi$, leaving this combination of observables dependant on the freeze-out temperature only, and constraining the region allowed in the SHM parameter space to a narrow band.   

It should be underlined that diagrams such as those in Fig. \ref{xifold} allow for a test of the physical validity of the SHM, since the SHM requires that {\em all} yields and fluctuations
be described by the {\em same} $\gamma_q$ and $T$.       If the measurement corresponding to the left panel of Fig \ref{xifold} is made, and the
result is {\em not} in the narrow band given in the figure, {\em or} if the temperature obtained in two of the three panels of Fig. \ref{xifold} is {\em not} the same, we can conclude that physics beyond the SHM plays a role.  Furthermore, the nature of the deviation gives a hint to its physical origin.

In particular, comparing fluctuations to directly detected resonances probes the interval between chemical and thermal freeze-out.
Consider, for example, the $\pi^+/\pi^-$ fluctuation.
The top and the bottom terms in this ratio are linked by a large correlation term due to the $\rho^0$ decay.   
This correlation probes the $\rho^0$ abundance at {\it chemical } freeze-out, since subsequent rescattering/regeneration does not alter the fact that the $\rho^0$ decay produced a $\pi^+$ and a $\pi^-$.
On the other hand, a direct measurement of the $\rho^0/\pi^-$ ratio through invariant mass reconstruction measures the $\rho^0$ abundance at {\em thermal} freeze-out, after all rescattering.
Hence, comparing the $\pi^+/\pi^-$ fluctuation   to the $\rho^0/\pi^-$ ratio provides a gauge for effect of the hadronic reinteraction period on particle abundances.   

The application of this method is outlined in the right panel of Fig. \ref{xifold}, where the measured $\rho^0$ abundance is compared to the $\pi^+/\pi^-$ fluctuation.  Again, the assumption of freeze-out with no reinteraction correlates the two observables to a narrow band Dependant on only the freeze-out temperature.  While a model-independent quantitative prediction of eventual deviations from the narrow
band is difficult to obtain a priori, one can infer qualitatively the likely origin of such deviations:  A shift down the $\rho^0/\pi^-$ axis would signal a long re-interacting phase which re-equilibrates the directly detectable $\rho^0$ to a lower ``thermal freeze-out'' temperature, suppressing their detectable yield but maintaining the correlation of the decay products.
A fall in yields together with a rise in fluctuations would be evidence of $\rho^0$ melting in-medium, while a rise in yields together with a fall in fluctuations would mean the $\rho^0$ abundance is augmented, presumably by an in-medium mass decrease.  A rise in both $\rho^0$ yield and fluctuation would be very problematic to explain in a model where statistical mechanics plays a role.

%%%%%%%%%%%%%%%%%%%%%%%%%%
%\begin{figure}[tb]
%\psfig{width=8.6cm,clip=,figure=pirho_fold.eps}
%\caption{\label{rhofold}
%A diagram correlating the abundance of the $\rho$ resonance with the fluctuations in the $\pi^+/\pi^-$ ratio.}
%\end{figure}
%%%%%%%%%%%%%%%%%%%%%

In conclusion, we have shown that the two statistical hadronization scenarios shown to be applicable at SPS and RHIC, one with a chemical freeze-out at 170 MeV and a long reinteraction
phase, the other with an explosive non-equilibrium transition from a high-entropy phase at 140 MeV, give definite and very different predictions for the interdependence of particle
yields and event-by-event fluctuations, allowing us to falsify either of these scenarios when both yields and fluctuations are taken into account.   We eagerly await more published data in both yields and fluctuations to determine weather the non-equilibrium
model is really capable of accounting for both yields and fluctuations in all light and strange hadrons produced in heavy ion collisions.

Work supported in part by grants from
the U.S. Department of Energy  (J.R. by DE-FG02-04ER41318)
the Natural Sciences and Engineering research
council of Canada, the Fonds Nature et Technologies of Quebec.
G. T. thanks the Tomlinson foundation  for support, and the QM2005 
organizers for financial support during the conference.
S.J.~thanks RIKEN BNL Center
and U.S. Department of Energy [DE-AC02-98CH10886] for
providing facilities essential for the completion of this work.\\

\end{document}